\documentclass[%
 aip,
 jcp,
 sd,%
 amsmath,amssymb,
preprint,%
]{revtex4-1}

\usepackage{graphicx}
\usepackage{dcolumn}
\usepackage{bm}

\begin{document}

\preprint{AIP/123-QED}

\title{Structure, hydrolysis and diffusion of aqueous vanadium ions from Car-Parrinello molecular dynamics}

\author{Zhen Jiang}
 \affiliation{Department of Chemical and Biomolecular Engineering, University of Nebraska-Lincoln, Lincoln, NE 68588, USA}
\author{Konstantin Klyukin}%
\affiliation{Department of Chemical and Biomolecular Engineering, University of Nebraska-Lincoln, Lincoln, NE 68588, USA}
\author{Vitaly Alexandrov}
\email{valexandrov2@unl.edu}
\affiliation{Department of Chemical and Biomolecular Engineering, University of Nebraska-Lincoln, Lincoln, NE 68588, USA}
\affiliation{Nebraska Center for Materials and Nanoscience, University of Nebraska-Lincoln, Lincoln, NE 68588, USA}

\date{\today}

\begin{abstract}
A molecular level understanding of the properties of electroactive vanadium species in aqueous solution is crucial for enhancing the performance of vanadium redox flow batteries (RFB). Here, we employ Car-Parrinello molecular dynamics (CPMD) simulations based on density functional theory to investigate the hydration structures, first hydrolysis reaction and diffusion of aqueous V$^{2+}$, V$^{3+}$, VO$^{2+}$, and VO$_2^+$ ions at 300 K. The results indicate that the first hydration shell of both V$^{2+}$ and V$^{3+}$ contains six water molecules, while VO$^{2+}$ is coordinated to five and VO$_2^+$ to three water ligands. The first acidity constants (p$K_\mathrm{a}$) estimated using metadynamics simulations are 2.47, 3.06 and 5.38 for aqueous V$^{3+}$, VO$_2^+$  and VO$^{2+}$, respectively, while V$^{2+}$ is predicted to be a fairly weak acid in aqueous solution with a p$K_\mathrm{a}$ value of 6.22. We also show that the presence of chloride ions in the first coordination sphere of the aqueous VO$_2^+$ ion has a significant impact on water hydrolysis leading to a much higher p$K_\mathrm{a}$ value of 4.8. This should result in a lower propensity of aqueous VO$_2^+$ for oxide precipitation reaction in agreement with experimental observations for chloride-based electrolyte solutions. The computed diffusion coefficients of vanadium species in water at room temperature are found to increase as V$^{3+}$ $<$ VO$_2^+$ $<$ VO$^{2+}$ $<$ V$^{2+}$ and thus correlate with the simulated hydrolysis constants, namely, the higher the p$K_\mathrm{a}$ value, the greater the diffusion coefficient.
\end{abstract}

\maketitle

\section{\label{sec:level1}INTRODUCTION}

There is a pressing demand to substitute fossil fuels with renewable energy sources such as solar and wind, but their intermittent and unpredictable nature calls for the development of advanced electrical energy storage technologies. In this regard, redox flow batteries (RFB) are one of the most promising solutions for large-scale grid energy storage that offer a number of advantages over their solid-state counterparts including the decoupling of power and energy, rapid response times, long cycling lifespan and the capability to store up to multimegawatt-hours of electrical energy.\cite{Weber2011,Alotto2014325,Grigorii2015}

Among various RFB, all-vanadium RFB offer some additional appealing properties such as  high electrochemical stability, reversibility and low cross-contamination by crossover between electrode compartments.\cite{Skyllas1986,Kear2012,Parasuraman201327}
The operation of all-vanadium RFB is based on the electrochemical reactions of the VO$^{2+}$/VO$_{2}^{+}$ redox couple in the catholyte and the V$^{2+}$/V$^{3+}$ in the anolyte, and the typical supporting electrolyte is a sulfuric acid solution. One of the major drawbacks of the current vanadium RFB systems is their low specific energy density ($<$30 Wh/kg), as compared to other rechargeable battery types such as lithium-ion batteries (up to 200 Wh/kg), at a narrow window of operating temperatures (10-40 $^{\circ}$C) due to various precipitation reactions.\cite{Huang2015} Increasing the energy density by simply making a more concentrated vanadium solution results in poor electrolyte stability and the formation of precipitates by all four vanadium cations that limits their aqueous concentrations\cite{Ding2013}. For instance, the stability of VO$_{2}^+$ in sulfuric acid is limited by V$_2$O$_5$ precipitation at high temperatures ($>$ 40 $^{\circ}$C), and although the VO$_{2}^+$ solubility can be increased by adding more sulfuric acid to the solution, the increase in acid concentration is limited by the stability of VO$^{2+}$ species that are present in the same electrolyte tank.\cite{Leung2012} Overall, the presence of all four electroactive vanadium species in the electrolyte is expected to lead to a very rich aqueous chemistry with various hydrolysis, polymerization and precipitation reactions, as demonstrated by the corresponding phase diagrams.\cite{Greenwood1998} Their equilibria depend strongly on the concentrations of both cationic and anionic species, pH, temperature and the charge/discharge state of the flow battery.\cite{Skyllas1999} Such a complex interplay between various factors makes the design of vanadium RFB electrolytes very challenging and requires a systematic approach in studying elementary reactions involved in these processes.

In the past years, there have been great efforts to improve the energy density of the vanadium RFB by modifying the chemistry of vanadium electrolyte solutions. For example, it was demonstrated that the use of a mixed sulfate-chloride electrolyte results in a 70\% increase in energy capacity over a broader operating temperature window as compared to the typical sulfate based electrolyte.\cite{Li2011394} Through a combination of the nuclear magnetic resonance (NMR) spectroscopy and density functional theory (DFT) calculations\cite{Li2011394,Vijayakumar20113669} it was shown that this improvement is most likely due to the formation of soluble VO$_2$Cl(H$_2$O)$_2$ species that impedes V$_2$O$_5$ precipitation. However, the use of halides in the electrolyte may lead to safety issues because of the formation of chlorine and bromine vapors at elevated temperatures. Therefore, some other studies focused on the use of safer organic and inorganic precipitation inhibitors to stabilize vanadium solutions.\cite{Roe2016,Chang2012}

One of the mechanisms underlying precipitation of metal oxides from the aqueous phase involves deprotonation of the first-shell water molecules and subsequent polymerization reactions via the formation of bridging oxygens or hydroxyls between the metal ions. In the case of vanadium electrolytes, it is believed that the formation of insoluble V$_2$O$_5$ occurs through this mechanism from VO$_2$(H$_2$O)$_3$$^+$ species present in solution.\cite{Li2011394} Therefore, it is indispensable to understand the thermodynamics and kinetics of the hydrolysis of the vanadium aqua species.

The transport of vanadium ions in the electrolyte and across the membrane is another key property of the vanadium RFB. The increase in concentration of vanadium ions in the electrolyte should lead to saturation and undesirable precipitation reactions. In addition, although the problem of cross-contamination is substantially diminished in the case of all-vanadium RFB due to the use of the same element for the redox-active couples, their transport during long-term use can be significant. In experiments it is often difficult to deconvolute the contributions from different vanadium species to the overall ionic transport, especially due to various possible accompanying chemical reactions between vanadium ions, anionic species in the electrolyte and membrane constituents.\cite{Luo2012,Knehr2012} Thus, it is important to examine transport properties and how they vary depending on the solution chemistry under well controlled conditions.

In the past, there were a few first-principles studies aiming to investigate the structural and thermodynamic properties of aqueous vanadium cations. While some investigations were limited to modeling hydrated vanadium species in the gas phase,\cite{Ralf1994,Vijayakumar20107709,Asmis2007} very few took into account the influence of aqueous solution under elevated temperatures.\cite{Sepehr201353,Sepehr2015} For example, the hydration structure of all four vanadium cations (V$^{2+}$, V$^{3+}$, VO$^{2+}$, and VO$_2^+$) was recently determined employing the gas-phase \textit{ab initio} calculations, whereas the thermodynamics of solvation was considered within the quasi-chemical theory of solvation.\cite{Sepehr201353} This study was able to provide basic information on the structural parameters of the most stable hydrous vanadium cations at zero temperature in the absence of aqueous solution. The effect of temperature and solvent on the structure and stability of solvated VO$_{2}^+$ was previously considered within the Car-Parrinello molecular dynamics (CPMD) methodology.\cite{Sadoc2007} Specifically, the study demonstrated that the penta-coordinated VO$_2$(H$_2$O$)_3^{+}$ complex should be more predominant than hexa-coordinated VO$_2$(H$_2$O)$_4^{+}$ complex in aqueous solution at room temperature, in agreement with the gas-phase calculations.\cite{Sepehr201353} Another CPMD study has focused on the evaluation of solvent effects on $^{51}$V NMR chemical shifts and also provided some useful information on the VO$_{2}^{+}$ speciation in aqueous solution and water dynamics around the vanadium center, albeit at the short timescale of 2 ps.\cite{Buhl2001}

In this paper, we present a systematic CPMD-based investigation of the hydration shell geometries, first hydrolysis constants (p$K_\mathrm{a}$) and diffusion coefficients of all four electroactive vanadium cations (V$^{2+}$, V$^{3+}$, VO$^{2+}$, and VO$_2^{+}$) in aqueous solution. The remainder of the paper is organized as follows. First, the computational methodology and details of the simulations are described. Second, the obtained results along with a discussion of the atomic structure, water dynamics, first hydrolysis reaction and diffusion of aqueous vanadium species are presented. At the end, main conclusions of the work are provided.

\section{COMPUTATIONAL DETAILS}
Car-Parrinello molecular dynamics (CPMD)\cite{CarParrinello1985} simulations were carried out for V$^{2+}$, V$^{3+}$, VO$^{2+}$ and VO$_2^+$ cations in aqueous solution in the canonical ensemble at 300 K employing the pseudo-potential plane-wave density functional theory (DFT) formalism as implemented in the NWChem code. \cite{Valiev20101477} Each cation was placed in a cubic box of length 13 {\AA} subject to periodic boundary conditions filled with 74 H$_{2}$O molecules to obtain the water density of approximately 1 g$/$cm$^3$. The charges on the vanadium cations were neutralized by a uniform negative charge background. Exchange and correlation were treated using the Perdew-Burke-Ernzerhof (PBE) functional\cite{Perdew1996}  within the generalized gradient approximation. To describe electron-ion interactions the norm-conserving pseudopotentials modified into a separable form according to Kleinman and Bylander were employed.\cite{Kleinman1982} Troullier-Martins pseudo-potentials\cite{Troullier1991} were used for vanadium and Hamann-type pseudo-potentials\cite{Hamann1979,Hamann1989} were used for oxygen and hydrogen. The kinetic energy cuttoffs of 60 and 120 Ry were applied to expand the Kohn-Sham electronic wave functions and charge density, respectively. Note that similar energy cutoffs were previously employed in CPMD simulations with the same pseudopotentials to examine hydration structures and stability of various cationic species in aqueous solutions,\cite{Sprik2001,Sadoc2007,Nichols2008} as well as for estimating p$K_\mathrm{a}$ values from \textit{ab initio} metadynamics.\cite{Tummanapelli2014}  The Brillouin zone was sampled with the ${\Gamma}$ point in all simulations.
The Nose-Hoover thermostat\cite{Nose1984,Hoover1985} was used to maintain the system at a temperature of 300 K and  the hydrogen atoms were replaced with deuterium to facilitate numerical integration. A fictitious electronic mass of 600 au and a simulation time step of $ {\delta t}$ = 5 au (0.121 fs) were set.  Each system was initially equilibrated during 3.6 ps using a QM/MM potential,\cite{Cauet2010} followed by a subsequent CPMD equilibration for 6 ps. The total production times were 21 ps for aqueous V$^{2+}$, V$^{3+}$, VO$^{2+}$ and 30 ps for aqueous VO$_2^+$. The collection time was longer in the case of VO$_2^+$ because the structure of its first hydration shell changed at about 6.5 ps during simulation, which will be discussed below. Configurations from the post-equilibration CPMD simulations used for further analysis were saved at time intervals of 10${\delta}$t.

To compute the free energy profiles of hydrolysis reaction ($\Delta F$) at 300 K for all four aqueous vanadium cations we employed \textit{ab initio} metadynamics technique.\cite{Micheletti2004,Laio2008}  To describe the H transfer from the water molecules in the first coordination sphere of vanadium ions we selected the following function $\xi(|\mathbf{r}_\mathrm{OH}|)$ for the coordination number as the collective variable (CV):

\begin{equation}
 \xi(|\mathbf{r}_\mathrm{OH}|) = \sum^{N_H}_{i=1}\frac{1}{1 + \exp[k(|\mathbf{r}_\mathrm{OH_i}| - r_\mathrm{cut})]}
 \label{eq1}
\end{equation}

Here, $N_{H}$ is the total number of protons, $k$ an arbitrary positive constant used to reproduce the equilibrium coordination number (set to 10 {\AA}$^{-1}$) and $r_\mathrm{cut}$ the O$-$H cutoff distance (set to 1.38 {\AA}). This choice of the collective variable to describe the first deprotonation reaction was previously shown to provide good results for other metal ions in aqueous solutions.\cite{Odoh2013} Metadynamics simulations at 300 K were performed using equilibrated geometries of aqueous vanadium ions picked from the last block of the CPMD production runs. The height and width of the repulsive Gaussian hills were set to 0.0001 au (0.063 kcal/mol) and 0.07, respectively. Gaussian hills were added to the potential every 100$\delta$t for V$^{3+}$(aq) and VO$_2^+$(aq) and  50$\delta$t for V$^{2+}$(aq) and VO$^{2+}$(aq). The time interval was shorter for V$^{2+}$(aq) and VO$^{2+}$(aq) since their deprotonation energies are expected to be greater than those for V$^{3+}$(aq) and VO$_2^+$(aq). Once the free energy profiles are simulated, the first hydrolysis constant (p$K_\mathrm{a}$) can be computed by using the standard relation
$\mathrm{p}K_\mathrm{a} = \Delta F/2.303 RT$. A more detailed discussion of the computational parameters employed in metadynamics simulations can be found in the Supporting Material.

\section{RESULTS AND DISCUSSION}
\subsection{Hydration Shell Geometry.}

\begin{figure}
\includegraphics[width=.4\textwidth]{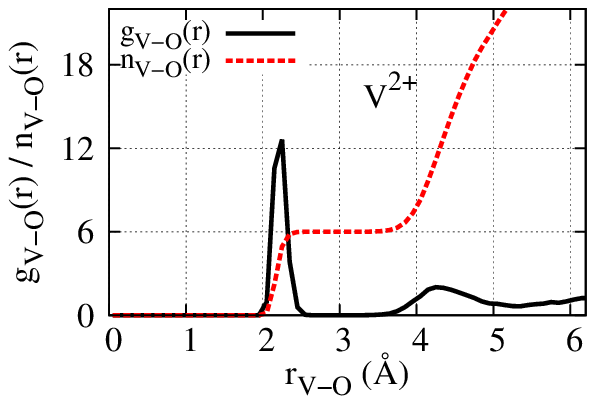}
\includegraphics[width=.4\textwidth]{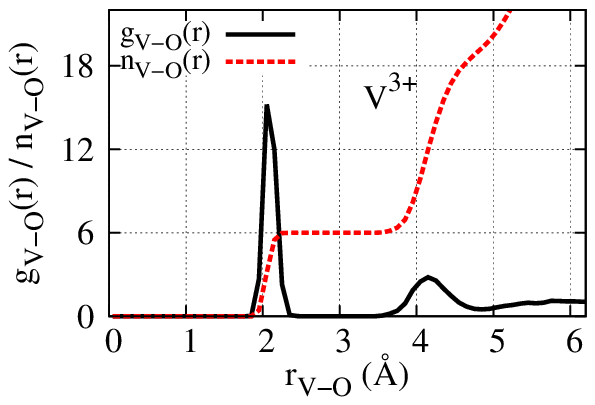}
\caption{Partial V$-$O radial distribution function (RDF) g$_\mathrm{{V-O}}$(r) and running coordination number n$_\mathrm{{V-O}}$(r) for aqueous V$^{2+}$ (left panel) and V$^{3+}$ (right panel).}
\label{fig1}
\end{figure}

\begin{table*}
\caption{Average distances R (in {\AA}) and angles O-V-O (in degrees) for the first hydration shell of V$^{2+}$ and V$^{3+}$ in water and their comparison with previous gas-phase calculations and experimental data. R$_\mathrm{{V-O}}$ stands for the distance between vanadium ions and oxygen atoms of the first-shell water molecules, R$_\mathrm{O-H}$ corresponds to the O$-$H bond length of water ligands, O$_{eq}$VO$_{ax}$ and O$_{ax}$VO$_{ax}$ are bond angles where "ax" denotes an axial and "eq" equatorial first-shell water molecule (see Figure \ref{fig2}).}
\begin{ruledtabular}
\begin{tabular}{ccccccc}
    {}& Method & $^a$N$_{H_2 O}$ & R$_{V-O}$ & R$_{O-H}$ & ${\angle}$O$_{eq}$VO$_{ax,eq}$ & ${\angle}$O$_{ax}$VO$_{ax}$ \\ \hline
    V$^{2+}$& this work & 6 & 2.23 & 0.97	& 89.75 & 170.91 \\
	{}& $^b$B3LYP$/$6-311+G$^{**}$\cite{Sepehr201353} & 6 & 2.21 & - & - & - \\
    {}& HF$/$$^c$ECP\cite{Ralf1994} & 6 & 2.20 & - & - & - \\
    {}& $^d$XRD\cite{Montgomery1967}& 6 & 2.15 &	- &	90.70 &	-\\
    {}& $^e$XRD\cite{Cotton1986} & 6 & 2.13 & - & 89.98 & 179.18 \\
  V$^{3+}$ & this work & 6 & 2.10 & 0.99 & 90.03 & 171.98 \\
    {}& B3LYP$/$6-311+G$^{**}$\cite{Sepehr201353} & 6 & 2.09 & - & - & - \\
    {}& HF$/$ECP\cite{Ralf1994} & 6 & 2.06 & - & - & - \\
    {}& $^f$XRD\cite{Cotton1984} & 6 & 2.00 & 1.00 & 91.74 & -\\
\end{tabular}
\end{ruledtabular}
\flushleft
\footnotesize $^a$The number of water molecules in the first coordination shell. $^b$Calculated for V(H$_2$O)$_6^{2+}$ with SMD in Gaussian09. $^c$The metal ions were described using the primitive basis sets optimized by Wachters, extended with two diffuse even-tempered $p$ functions and one $d$ function. $^d$Detected from (NH$_4$)$_2$V(SO$_4$)$_2$$\cdot$6H$_2$O using x-ray diffraction (XRD). $^e$Detected from VSO$_4$$\cdot$6H$_2$O. $^f$Detected from V(H$_2$O)$_6$(H$_5$O$_2$)(CF$_3$SO$_3$)$_4$.
\label{tab1}
\end{table*}

\begin{figure}
  \includegraphics[width=.3\textwidth]{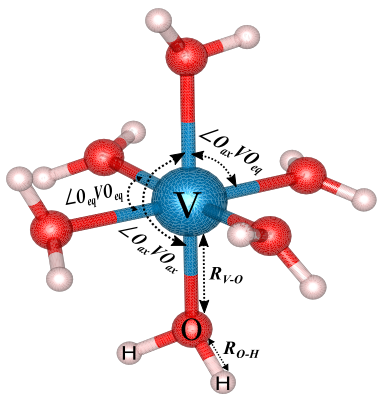}
  \caption{Snapshot from the CPMD trajectory of the first coordination sphere of aqueous V$^{2+}$ and V$^{3+}$.}
  \label{fig2}
\end{figure}

\textbf{V$^{2+}$(aq) and V$^{3+}$(aq).} We start by analyzing the atomic structure of vanadium ions in aqueous solution based on our CPMD results and Figure \ref{fig1} shows the radial distribution function (RDF) and running coordination numbers of aqueous V$^{2+}$ and V$^{3+}$ ions. For both ions two distinct peaks in the RDF can be identified. The well-defined first peak is in the 2.0 $-$ 2.5 {\AA} range for V$^{2+}$ and in the 1.9 $-$ 2.3 {\AA} range for V$^{3+}$, corresponding to the first hydration shell. It is seen from the running coordination numbers that the first hydration shell of both V$^{2+}$ and V$^{3+}$ has six water molecules, in agreement with previous  experimental\cite{Eldik1992} and \textit{ab initio} studies.\cite{Sepehr201353,Ralf1994} The second shell peak is characterized by a broader distribution between 3.7  and 5.1 {\AA} with an average V$-$O distance of 4.3 {\AA} for V$^{2+}$(aq) and 4.2 {\AA}  for V$^{3+}$(aq). Table \ref{tab1} lists the average structural parameters for the first hydration shell along with the data from previous studies, while Figure \ref{fig2} serves as a visual representation of the first shell structure for both V$^{2+}$ and V$^{3+}$ ions. The first shell is characterized by rather regular octahedral geometry around the vanadium centers with four water molecules in equatorial plane and two water molecules in axial direction. As seen from Table \ref{tab1}, the average V$-$O distance is by about 0.1 {\AA} shorter for V$^{3+}$(aq) reflecting its larger ionic size and stronger electrostatic attraction as compared to V$^{2+}$. The corresponding ${\angle}$OVO bond angles are found to be similar for V$^{2+}$ and V$^{3+}$. Overall, both the V$-$O bond lengths and ${\angle}$OVO bond angles are very similar to those obtained in previous computational studies\cite{Ralf1994,Sepehr201353} and agree well with available experimental data.\cite{Montgomery1967,Cotton1986,Cotton1984}


\begin{figure}
  \centering
  \includegraphics{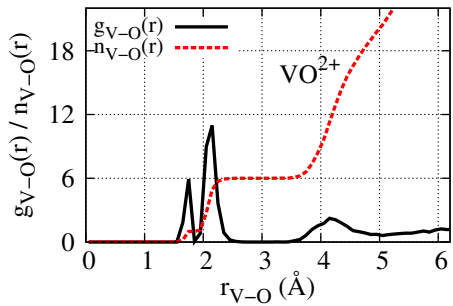}
  \caption{Partial V$-$O radial distribution function (RDF) g$_\mathrm{V-O}$(r) and running coordination number n$_\mathrm{V-O}$(r) for aqueous VO$^{2+}$.}
\label{fig3}
\end{figure}

\begin{table*}
\caption{The same as in Table \ref{tab1} for aqueous VO$^{2+}$ (see Figure \ref{fig4} for visual representation).}
\begin{ruledtabular}
\begin{tabular}{cccccccc}
    {}& Method & N$_{H_2 O}$ & R$_{V-O_{ax}}$ & $^a$R$_{V-O_{ax}^w}$ & R$_{V-O_{eq}}$ & ${\angle}$O$_{ax}$VO$_{eq}$ & ${\angle}$O$_{ax}$VO$_{ax}^w$ \\ \hline
    VO$^{2+}$& this work  & $5$ & $1.72$ & $2.24$	& $2.11$ & $97.40$ & $169.61$ \\
	{}& B3LYP$/$6-311+G$^{**}$\cite{Sepehr201353} & $5$ & 1.57 & 2.36 & 2.12 & - & - \\
    {}& $^b$DFT/QZ4P\cite{Vijayakumar20107709} & $5$ & $1.57$ & 2.36 & 2.12 & -& - \\
    {}& $^c$XRD\cite{Ballhausen1968}& $5$ & $1.59$ &	2.22 &	$2.04$ & 98.00 & 174.20 \\
    {}& $^d$LAXS\cite{Joanna2012} & $5$ & $1.62$ & $2.20$ & $2.03$ & - & -  \\
\end{tabular}
\end{ruledtabular}
\flushleft
\footnotesize $^a$O$_{ax}^w$ denotes the axial O atom of H$_2$O.  $^b$Calculated for V(H$_2$O)$_6^{3+}$ using QZ4P basis set (quad Z, 4 polarization functions, all electron) in ADF package. $^c$Detected from VOSO$_4$$\cdot$5H$_2$O. $^d$Detected from VOClO$_4$ in water with concentration of 1.228 mol$/$dm$^3$ using large angle x-ray scattering (LAXS).
\label{tab2}
\end{table*}

\begin{figure}[h]
  \centering
  \includegraphics[width=.3\textwidth]{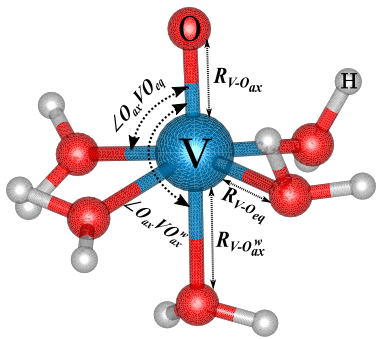}
  \caption{Snapshot from the CPMD trajectory of the first coordination sphere of aqueous VO$^{2+}$.}
\label{fig4}
\end{figure}

\textbf{VO$^{2+}$(aq).} The RDF and running coordination number of VO$^{2+}$(aq) are shown in Figure \ref{fig3}.
The first lower peak in the RDF curve is in the range of 1.6$-$1.8 {\AA} and corresponds
to the O atom of the VO$^{2+}$ cation, while the second taller peak between 1.8$-$2.4 {\AA} is associated with the oxygen atoms from the first hydration shell. Therefore, based on the n$_{V-O}$(r) curve there are five water molecules in the first hydration shell of VO$^{2+}$(aq). The average first-shell structural parameters along with some literature data are listed in Table \ref{tab2}, while the visual representation of the first hydration geometry with the corresponding definitions is given in Figure \ref{fig4}. The average bond length between vanadium and oxygen atom of the VO$^{2+}$ cation (the maximum of the first peak in the RDF at 1.72 {\AA}) is by about 0.5 and 0.4 {\AA} shorter than the bond distances between V$^{2+}$ and V$^{3+}$ ions and the oxygen from the first-shell water molecules, respectively. The geometry of the aqua-VO$^{2+}$ complex is a slightly more distorted octahedron than the ones observed for V$^{2+}$ and V$^{3+}$ with a larger average ${\angle}$O$_{ax}$VO$_{eq}$ of 97.4$^{\circ}$.

The longest bond distance of about  2.24 {\AA} is found to be between the vanadium and axial water oxygen (R$_{V-O^w_{ax}}$). The average V$-$O bond length in the equatorial plane (R$_{V-O_{eq}}$) is 2.11 {\AA} being very close to that of V$^{3+}$(aq). The values of the V$-$O$^w$$_{eq}$ and V$-$O$_{eq}$ bond distances reflect the stronger attraction force of VO$^{2+}$ to the five coordinated water molecules than that of V$^{2+}$, whereas it is weaker than for V$^{3+}$.


\begin{figure}[h]\centering
\includegraphics[width=.4\textwidth]{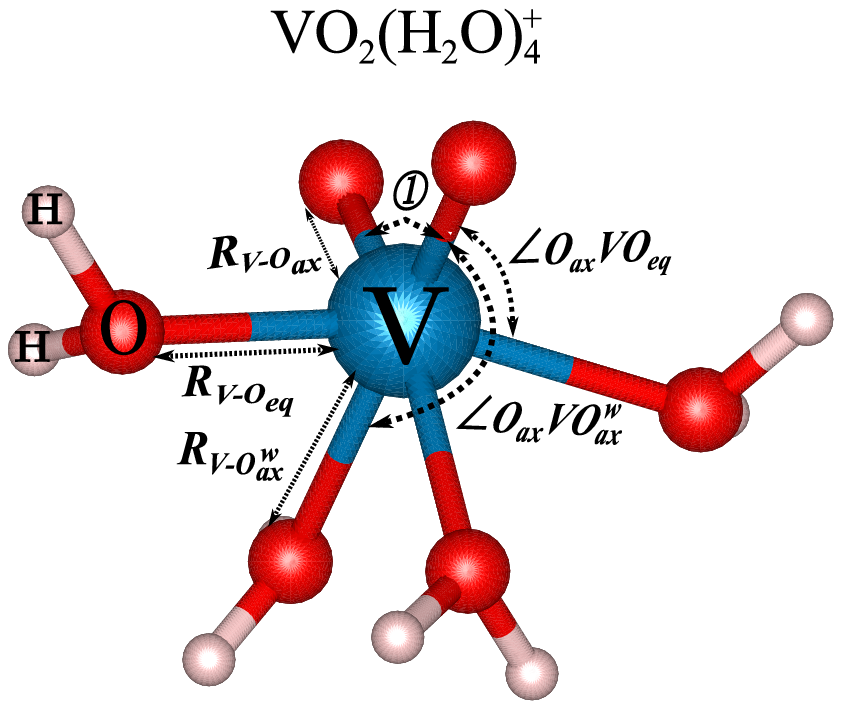}
\includegraphics[width=.4\textwidth]{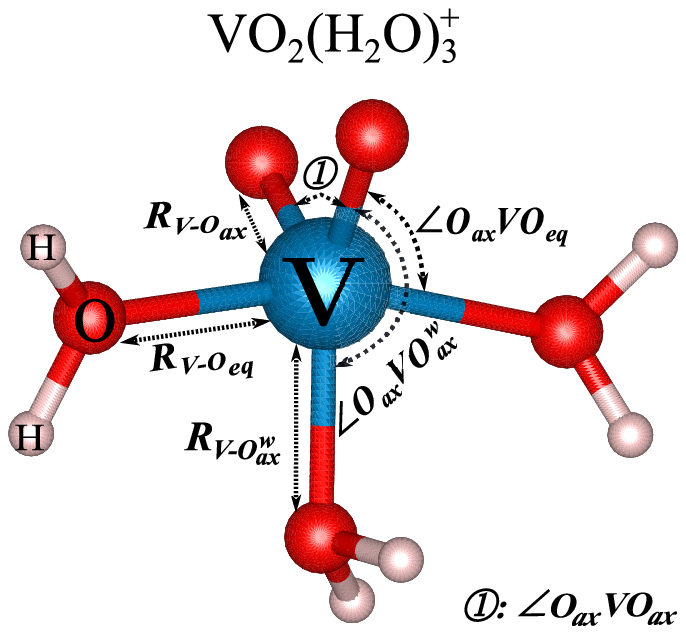}
\caption{CPMD trajectory snapshots of the first coordination sphere of aqueous VO$_2$(H$_2$O)$_4^{+}$ (left panel) and VO$_2$(H$_2$O)$_3^{+}$ (right panel).}
\label{fig5}
\end{figure}

\begin{figure}[h]
\centering
\includegraphics[width=.4\textwidth]{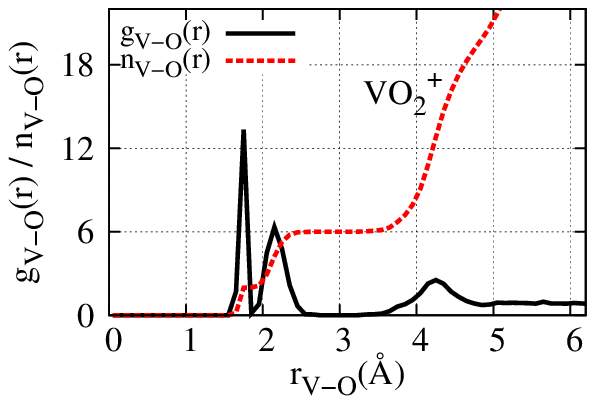}
\includegraphics[width=.4\textwidth]{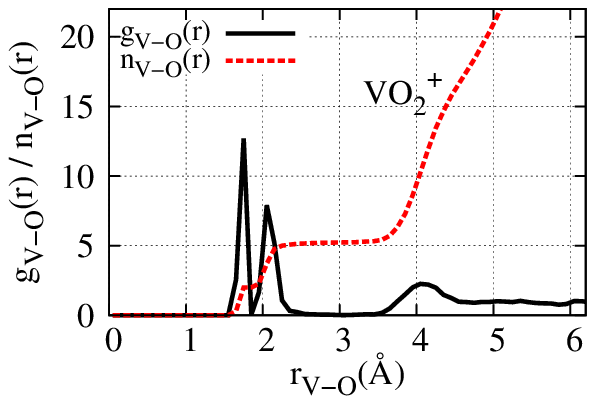}
\caption{Partial V$-$O radial distribution function (RDF) g$_\mathrm{V-O}$(r) and running coordination number n$_\mathrm{V-O}$(r) for two stable states of VO$_2$$^{+}$(aq): VO$_2$(H$_2$O)$_4^{+}$ (left panel) and VO$_2$(H$_2$O)$_3^{+}$ (right panel) which geometries are shown in Figure \ref{fig5}.}
\label{fig6}
\end{figure}

\begin{figure}[h]
  \centering
  \includegraphics[width=.4\textwidth]{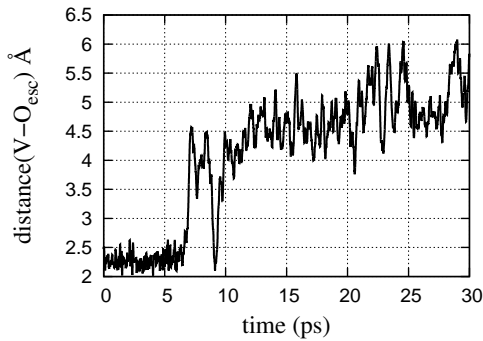}
  \caption{Variation of the bond distance between V and O$_\mathrm{esc}$, where O$_\mathrm{esc}$ atom belongs to the labile water molecule escaping from the first to the second shell of VO$_2$(H$_2$O)$_4^{+}$(aq) during CPMD simulation.}
  \label{fig7}
\end{figure}

\begin{figure}[h]\centering
\includegraphics[width=.4\textwidth]{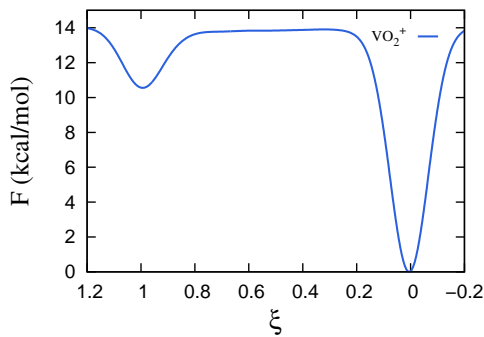}
\caption{Free energy profile of the reaction VO$_2$(H$_2$O)$_4^{+}$ ${\rightleftharpoons}$ VO$_2$(H$_2$O)$_3^{+}$ + H$_2$O in aqueous solution, where $\xi=1$ corresponds to the reactant VO$_2$(H$_2$O)$_4^{+}$ and $\xi=0$ to the product VO$_2$(H$_2$O)$_3^{+}$ + H$_2$O. }
\label{fig8}
\end{figure}

\textbf{VO$_2^{+}$(aq)} In the case of VO$_2^{+}$ ion we observe two stable hydration shell structures during CPMD simulation (Figure \ref{fig5}). The starting geometry corresponds to VO$_2$(H$_2$O)$_4^{+}$ complex with four water ligands in the first shell as seen from the RDF and running coordination number (see left panel of Figure \ref{fig6}). After approximately 6.5 ps one water molecule is expelled from the first to the second coordination sphere of VO$_2^{+}$ as shown by the variation of the corresponding V$-$O distance as a function of time in Figure \ref{fig7} to form a more stable trigonal bipyramidal VO$_2$(H$_2$O)$_3^{+}$ complex. Previously, the trigonal bipyramidal structure was estimated to be about 8.6 kcal$/$mol more stable than the hexagonal aqua complex at zero temperature using the gas-phase calculations at the B3LYP$/$6-311+G$^{**}$ level.\cite{Sepehr201353} To get some insight into the energetics of this chemical transformation between two complexes in solution at 300 K, we use metadynamics simulations to find that the VO$_2$(H$_2$O)$_3^{+}$ species is indeed by about 10.6 kcal/mol more stable with the energy barrier of its formation from VO$_2$(H$_2$O)$_4^{+}$ of only 3.2 kcal/mol (Figure \ref{fig8}).  Thus, our results support that the tri-aqua VO$_2^{+}$ complex should be dominant VO$_2^{+}$ species in the aqueous phase at 300 K.

\begin{table*}
\caption{The same as in Table \ref{tab1} for two stable aqua complexes of VO$_2^{+}$ (see Figure \ref{fig5} for visual representations).}
\begin{ruledtabular}
\begin{tabular}{ccccccccc}
    {}& Method$^{ref.}$ & N$_{H_2 O}$ & R$_{V-O_{ax}}$ & R$_{V-O_{ax}^w}$ & R$_{V-O_{eq}}$ & ${\angle}$O$_{ax}$VO$_{ax}$ & ${\angle}$O$_{ax}$VO$_{eq}$ & ${\angle}$O$_{ax}$VO$_{ax}^w$ \\ \hline

    VO$_2$(H$_2$O)$_4^{+}$& this work & 4 & $1.73$ & $2.27$	& $2.08$ & $105.54$ & $93.05$ & $162.41$\\
	{}& B3LYP$/$6-311+G$^{**}$\cite{Sepehr201353} & 4 & $1.61$ & $2.32$ & $2.10$ & - & - & - \\
    {}& $^a$BP86/6-31G$^{*}$/all-electron\cite{Buhl2001} & 4 & $1.61$ & $2.28$ & $2.13$ & - & - & - \\
    {}& $^b$LAXS\cite{Joanna2012}& 4 & $1.63$ &	$2.23$ & $2.01$ & - & - & - \\
    VO$_2$(H$_2$O)$_3^{+}$& this work  & 3 & $1.72$ & $2.12$	& $2.05$ & $108.77$ & $97.96$ & $^c$143.30, 106.31\\
	{}& B3LYP$/$6-311+G$^{**}$\cite{Sepehr201353} & 3 & $1.60$ & $2.11$ & - & - & - & - \\
    {}& BP86/6-31G$^{*}$/all-electron\cite{Buhl2001} & 3 & $1.61$ & $2.10$ & $2.13$ & - & - & - \\
    {}& $^d$B3LYP/QZ4P\cite{Vijayakumar20113669}& 3 & $1.64$ &	$2.13$ & $2.10$ & - & - & - \\
\end{tabular}
\end{ruledtabular}
\flushleft
\footnotesize $^a$All-electron Wachters' basis set with two diffuse $d$ and one diffuse $p$ functions for vanadium and 6-31G$^{*}$ basis set for all other atoms.\cite{Buhl2001} $^b$Detected in sulfuric acid solution with concentration of 1.833 mol$/$dm$^3$. $^c$142.79$^{\circ}$ corresponds to the O$_{ax}$ atom above the paper plane and 108.21$^{\circ}$ to the O$_{ax}$ below the paper plane. $^d$Conductor-like screening model (COSMO) was employed for treating aqueous solution.
\label{tab3}
\end{table*}

Table \ref{tab3} lists the average structural parameters for VO$_2$(H$_2$O)$_4^{+}$ and VO$_2$(H$_2$O)$_3^{+}$ along with the available data from previous studies, while Figure \ref{fig5} provides representative snapshots from the CPMD trajectories for the two first-shell geometries.
We find that  the average values of R$_{V-O_{eq}}$ and R$_{V-O_{ax}}$ for VO$_2$(H$_2$O)$_4^{+}$ are effectively the same as those of VO$_2$(H$_2$O)$_3^{+}$. As for the ${\angle}$OVO bond angles in VO$_2$(H$_2$O)$_3^{+}$, two different values (143.30$^{\circ}$ and 106.31$^{\circ}$) are reported to show the average ${\angle}$O$_{ax}$VO$_{ax}^w$ corresponding to the O$_{ax}$ atom above and below the paper plane, respectively. Thus, combining these two values with  108.77$^{\circ}$ of ${\angle}$O$_{ax}$VO$_{ax}$, the sum is equal to 358.38$^{\circ}$ indicating that all three axial water oxygen atoms and central vanadium
cation shown in Figure \ref{fig5} are in the same plane. The average ${\angle}$O$_{ax}$VO$_{eq}$ is equal to 97.96$^{\circ}$ showing that the VO$_2$(H$_2$O)$_3^{+}$ structure is not a perfect trigonal bipyramidal.

\subsection{Metadynamics Simulations of First Acidity Constant}

Understanding the hydrolysis reactions of vanadium species in the aqueous phase is of great significance for improving the stability of supporting electrolytes in vanadium RFB. For example, it is believed that the solvated VO$_{2}^+$ can be converted to insoluble V$_2$O$_5$ according to the following deprotonation and condensation reactions:\cite{Vijayakumar20113669}
\begin{equation}
\mathrm{[VO_{2}(H_2O)_3]^+ {\rightleftharpoons} VO(OH)_3 + H_3O^+}
\end{equation}
\begin{equation}
\mathrm{2VO(OH)_3 {\rightleftharpoons} V_2O_5 \cdot 3H_2O}
\end{equation}

Depending on solution chemistry, pH and thermodynamic conditions, the deprotonated species can form new vanadium complexes and precipitates that will limit stability and redox characteristics of vanadium ions in aqueous solution.\cite{Li2011394} Therefore, it is important to understand the propensity of different vanadium species for deprotonation under ambient conditions.  Here, we analyze the mechanism and kinetics of the first hydrolysis reaction across all four vanadium species using the metadynamics simulations. For example, the first deprotonation reaction of [VO$_2$(H$_2$O)$_3$]$^+$ complex in water can be described as:
\begin{equation}
\mathrm{[VO_2(H_2O)_3]^+ + H_2O {\rightleftharpoons} VO_2(H_2O)_2(OH) +H_3O^+}
\end{equation}

First of all, we note that the proton transfer from the first to the second hydration shell did not result in any significant transformation of the vanadium complexes except some reasonable changes in interatomic distances and angles. The vanadium ions did not change their coordination number as no water molecule departed from the first shell during hydrolysis. It is expected that the higher charged V$^{3+}$(aq) will undergo the hydrolysis reaction easier being a much stronger acid in aqueous solution than other vanadium species. The measured p$K_\mathrm{a}$ value of V$^{3+}$(aq) is indeed smaller than those of the other cations and our calculations support this observation (see Table \ref{tab4}).

\begin{figure}
  \includegraphics{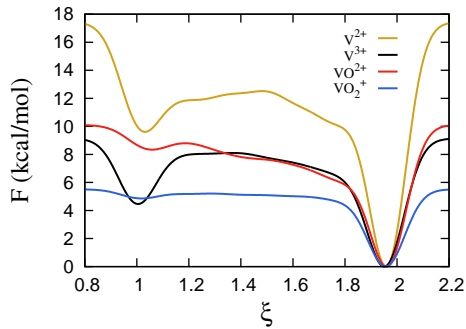}
  \caption{Free energy profiles of the first deprotonation reaction for aqueous V$^{2+}$, V$^{3+}$, VO$^{2+}$ and VO$_2^{+}$. $\xi$ $=$ 1 corresponds to the product and $\xi$ $=$ 2 to the reactant state.}
  \label{fig9}
\end{figure}

\begin{table}
\caption{Entropy corrected free energies ${\Delta}$F (in kcal/mol) and p$K_\mathrm{a}$ values of aqueous V$^{2+}$, V$^{3+}$, VO$^{2+}$ and VO$_2^{+}$.}
\begin{ruledtabular}
\begin{tabular}{cccc}
    {}& ${\Delta}$F  & p$K_\mathrm{a}$ & p$K_\mathrm{a}$ (exp.) \\ \hline
    V$^{2+}$& $8.54$ & $6.22$	& -  \\
	V$^{3+}$& 	$3.39$ & $2.47$	& 2.8 \cite{walker2012}, $2.9$\cite{Yatsimirskir1960,McCleverty2004}\\
    VO$^{2+}$&  $7.38$ & $5.38$ & 5.35\cite{Francavilla1975}, 5.66\cite{Henry1973}, 5.67\cite{McCleverty2004}, 6.0\cite{Rossotti1955}\\
    VO$_2^{+}$&  $4.20$ & $3.06$	& 3.2\cite{McCleverty2004,Akio1979}\\
\end{tabular}
\end{ruledtabular}
\label{tab4}
\end{table}

\begin{figure}
  \includegraphics{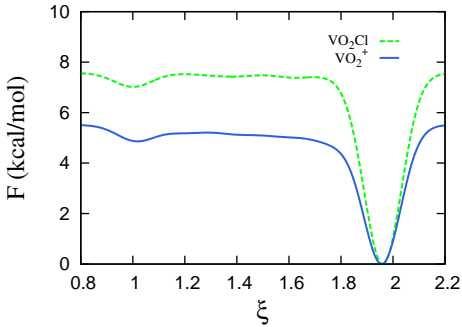}
  \caption{Free energy profiles of the first deprotonation reaction for aqueous VO$_{2}$Cl(H$_2$O)$_2$ versus [VO$_{2}$(H$_2$O)$_3$]$^+$ species.}
  \label{fig10}
\end{figure}

Figure \ref{fig9} shows the simulated free energy profiles of the deprotonation reaction as a function of the coordination number $\xi$ of a specific first-shell water oxygen atom, as defined by Equation \ref{eq1}. In this figure $\xi \approx 2$ corresponds to the reactant, taken as the zero reference point, and $\xi \approx 1$ corresponds to the product state with one proton departed from the first-shell water molecule to form a hydroxyl attached to a vanadium ion. Thus, the free energy difference between these two states can be directly related to p$K_\mathrm{a}$ (see Table \ref{tab4}). It is seen that V$^{3+}$(aq) is the strongest acid with $ \Delta F = F(\xi = 1.01) - F(\xi = 1.95) = 4.46$ kcal/mol. Since any of the six water molecules from the first shell could participate in the hydrolysis reaction an entropic energy correction of $-TS = -300k_b\ln 6 = -1.07$ kcal/mol must be added to the free energy. This gives the final estimate of $\Delta$F to be 3.39 kcal/mol yielding a p$K_\mathrm{a}$ value of 2.47. The VO$_{2}^+$(aq) is characterized by a slightly larger $\Delta$F value of 4.86 kcal/mol. Since we found that  the VO$_{2}^+$ ion prefers to be coordinated with three water molecules, it has the smallest entropic correction to the free energy among all vanadium ions of only $-k_bT\ln 3$ = -0.66 kcal/mol resulting in a p$K_\mathrm{a}$ value of 3.06. The p$K_\mathrm{a}$ value of VO$^{2+}$(aq) is calculated to be 5.38, in very good agreement with the 5.3$-$6.0 range in experimental values.\cite{Francavilla1975,Rossotti1955,Henry1973,McCleverty2004} The calculated value is larger than that of the  V$^{3+}$ due to the presence of oxygen ion in VO$^{2+}$(aq) that makes the effective charge on the vanadium center smaller thus strengthening the O$-$H bond of the dissociating water molecule. The p$K_\mathrm{a}$ value of V$^{2+}$(aq) is predicted to be the largest being 6.22. The less effective charge transfer from the first-shell water molecule to the V$^{2+}$ ion should make the release of proton less favorable, which is consistent with the longest average bond length between vanadium and oxygen atom of the first-shell water molecule observed for V$^{2+}$ among all vanadium ions.  The low propensity of V$^{2+}$(aq) for hydrolysis agrees well with the experimental observations of good stability of V$^{2+}$ at high V$^{2+}$ concentrations.\cite{Li2011394} Overall, our estimated p$K_\mathrm{a}$ values are a little smaller than the experimental values showing a reasonable agreement given the approximations made in our simulations.

We next examine the influence of chloride ions from the electrolyte solution on the deprotonation reaction of aqueous VO$_2^{+}$. It was previously demonstrated experimentally that the use of an aqueous electrolyte containing a mixture of H$_{2}$SO$_{4}$ and HCl acids can considerably inhibit the rate of V(V) precipitation reaction leading to a higher thermal stability of V(V) species in mixed-acid electrolyte solutions.\cite{Li2011394} This effect was attributed to the presence of chloride ions in the first shell of the aqueous VO$_2^{+}$ ion that should diminish the water deprotonation reaction. A recent \textit{ab initio} study has corroborated that the probability of the formation of VO$_{2}$Cl(H$_2$O)$_2$ species is very high and they are energetically favored over [VO$_{2}$(H$_2$O)$_3$]$^+$ complexes in a mixed chloride-sulfate solution.\cite{Bon2016} To quantitatively estimate the propensity of the VO$_{2}$Cl(H$_2$O)$_2$ complex for deprotonation reaction, we perform metadynamics simulations of the first acidity constant of an aqueous solution containing VO$_{2}$Cl(H$_2$O)$_2$ species. To this end, we first carry out additional 2.4 ps CPMD equilibration starting from the last CPMD block for VO$_2^{+}$(aq) with one water molecule substituted by a chloride ion.  Figure \ref{fig10} shows the free energy profiles for the chloride-free and chloride-containing species and the pK$_{a}$ value of 4.8 estimated for VO$_{2}$Cl(H$_2$O)$_2$ is indeed much greater than that of [VO$_{2}$(H$_2$O)$_3$]$^+$ (3.06) thus supporting experimental findings for chloride-containing electrolyte solutions.\cite{Li2011394}

\subsection{Diffusion Coefficients}

The vanadium ion migration is central to the operation of vanadium RFB. It is believed that the crossover of vanadium ions across the membrane may play a considerable role in causing the capacity loss  over long-term charge-discharge cycling.\cite{Sun2010890} Moreover, the increase in concentration of vanadium ions due to their net transfer may lead to undesirable precipitation reactions. Therefore, it is of great importance to provide detailed vanadium ion transport information. Here, we estimate the diffusion coefficients of all four vanadium ions in water based on the previously generated CPMD trajectories. Diffusion coefficients are calculated from a linear fit to the mean square displacement over time according to the relation
\begin{equation}
6Dt = <| \mathbf{r}(t) - \mathbf{r}(0) |^2>
\end{equation}
where $\mathbf{r}$($t$) is the position of the vanadium center at time $t$ and $D$ is the diffusion coefficient. The time interval used for the fit was 20 ps for each ion, but we note that in the case of VO$_2^{+}$ a more distant time interval (starting from $t$ = 7.5 ps) was chosen as it corresponds to the more energetically favorable complex with three water ligands in the first coordination shell. We also note that our evaluation of the diffusion coefficients serves only as a very approximate measure due to short simulation times, however, similar time scales were previously used to estimate diffusion coefficients at room temperature.\cite{Allesch2004} The details of the fits along with the corresponding plots are provided in the Supporting Material, while the computed diffusion coefficients are presented in Table \ref{tab5} along with experimental estimates.

\begin{table}
\caption{Calculated diffusion coefficients along with available experimental data for aqueous V$^{2+}$, V$^{3+}$, VO$^{2+}$ and VO$_2^{+}$ (in cm$^2/$s$\times$10$^{-6}$).}
\label{tab5}
\begin{ruledtabular}
\begin{tabular}{ccccc}
    {}& V$^{2+}$ & V$^{3+}$ & VO$^{2+}$ & VO$_2^{+}$ \\ \hline
    this work& 6.85 & 0.95 & 4.22 & 4.03 \\
    $^a$exp.\cite{Kim2001104}& 5.89 & - & - & - \\
    $^b$exp.\cite{Gattrell2004}& - & - &	(3.00 ${\pm}$ 0.30) & 2.00  \\
    $^c$exp.\cite{Oriji2005321}& 1.10 & 0.57 & 1.00 & 1.00 \\
\end{tabular}
\end{ruledtabular}
\flushleft
\footnotesize $^a$The value obtained by chronocoulometric method in acidic H$_2$O $+$ DMF mixed solutions. $^b$The values obtained in 1 mol/L sulfuric acid solution. $^c$The values obtained by cyclic voltammetry and sand's equation  in 5 mol/L sulfuric acid solution.
\label{tab5}
\end{table}
As it is seen from Table \ref{tab5}, our calculated diffusion coefficients for the four vanadium species in water follow the trend V$^{3+}$ $<$ VO$_2^{+}$ $<$ VO$^{2+}$ $<$ V$^{2+}$. The available experimental values are rather scattered due to various experimental conditions and techniques used to estimate the diffusion coefficients. It is interesting to note that here we observe the same trend as we found for the first deprotonation reaction (p$K_\mathrm{a}$) indicating that the strongest acid, V$^{3+}$(aq), exhibits the slowest diffusion, whereas the weakest acid, V$^{2+}$(aq), is characterized by the most rapid diffusion. This correlation can be rationalized in terms of the weak interaction between V$^{2+}$ and the surrounding water molecules leading to the largest p$K_\mathrm{a}$ value among all vanadium species and, as a result, to the fastest diffusion.

\section{CONCLUSIONS}
Properties of the V$^{2+}$/V$^{3+}$ and VO$^{2+}$/VO$_2^+$ redox couples play a central role in all-vanadium redox flow batteries. Here, we employed Car-Parrinello molecular dynamics simulations to investigate the atomic structure, first hydrolysis reaction and diffusion of all four vanadium ions in water at 300 K.
Our simulations show that both V$^{2+}$ and V$^{3+}$ cations exist as stable hexa-aqua octahedral complexes in solution, whereas VO$^{2+}$ is found to have five and VO$_2^+$ three water ligands in the first hydration shell. The octahedral VO$_2$(H$_2$O)$_4^+$ complex was found to be by about 10.6 kcal/mol less stable than the trigonal bipyramidal VO$_2$(H$_2$O)$_3^+$. The average distance between the vanadium ion and the oxygen atom from the first-shell water molecules R$_{V-O}$ changes as V$^{3+}$(aq) $\approx$ VO$_2^{+}$(aq) $<$ VO$^{2+}$(aq) $<$ V$^{2+}$(aq).

The free energy profiles of the first hydrolysis reaction obtained from Car-Parrinello metadynamics simulations were used to estimate the first acidity constant (p$K_\mathrm{a}$) across all aqueous vanadium species. During the proton transfer from a coordinated water molecule to a second-shell water molecule all water molecules  remain in the first coordination sphere of the vanadium center. Our simulated p$K_\mathrm{a}$ values are found to be 2.47, 3.06 and 5.38 for aqueous V$^{3+}$, VO$^{2+}$ and VO$_2^+$, respectively, being in good agreement with available experimental data. The aqueous V$^{2+}$ is predicted to be a fairly weak acid with a p$K_\mathrm{a}$ value of 6.22. The high p$K_\mathrm{a}$ value found for V$^{2+}$(aq) reflects its low propensity for hydrolysis and precipitation reactions and this is consistent with experimental observations of good stability of V$^{2+}$ at high V$^{2+}$ concentrations. We also show that the complexation of aqueous VO$_2^{+}$ with a chloride ion leads to a much higher p$K_\mathrm{a}$ value of 4.8 and thus should substantially inhibit the water deprotonation reaction and precipitation of the V(V) oxide in full agreement with experimental findings for vanadium ions in mixed sulfate-chloride solutions.

The computed diffusion coefficients of  V$^{3+}$, VO$_2^+$, VO$^{2+}$ and V$^{2+}$ in water at 300 K are 0.95, 4.03, 4.22 and 6.85 $\times$ 10$^{-6}$ cm$^2$/s, respectively, and thus follow the same trend as the first acidity constants, V$^{3+}$(aq) $<$ VO$_2^{+}$(aq) $<$ VO$^{2+}$(aq) $<$ V$^{2+}$(aq), with the strongest acid exhibiting slowest diffusion.

\section{Supplementary Material}
See supplementary material for the details on \textit{ab initio} metadynamics simulations including the choice of the number of Gaussian functions used to obtain free-energy profiles of the deprotonation reaction (Section I). Section II provides the details on the estimation of the diffusion coefficients along with the plots of the mean square displacement as a function of time (Figure S6).

\begin{acknowledgements}

V.A. gratefully acknowledges financial support from the startup package and the Layman seed grant provided by the University of Nebraska Foundation. The Holland Computing Center at the University of Nebraska-Lincoln is acknowledged for providing computational resources.

\end{acknowledgements}


\bibliography{aipsamp}
\end{document}